# Multi-scale Community Detection using Stability Optimisation within Greedy Algorithms

Erwan Le Martelot

Department of Computing

Imperial College London

London SW7 2AZ, United Kingdom

e.le-martelot@imperial.ac.uk

Chris Hankin

Department of Computing

Imperial College London

London SW7 2AZ, United Kingdom

c.hankin@imperial.ac.uk

July 9, 2018

**Abstract**

Many real systems can be represented as networks whose analysis can be very informative regarding the original system's organisation. In the past decade *community detection* received a lot of attention and is now an active field of research. Recently *stability* was introduced as a new measure for partition quality. This work investigates stability as an optimisation criterion that exploits a Markov process view of networks to enable multi-scale community detection. Several heuristics and variations of an algorithm optimising stability are presented as well as an application to overlapping communities. Experiments show that the method enables accurate multi-scale network analysis.

## 1 Introduction

In various fields such as biology, sociology, engineering and beyond, systems are commonly represented as graphs, or networks (e.g. protein networks, social networks, web). In the past decade the field of community detection attracted a lot of interest considering community structures as important features of real-world networks [10]. Commonly, given a network of any kind, looking for communities refers to finding groups of nodes that are more densely connected internally than with the rest of the network. The concept considers the inhomogeneity within the connections between nodes to derive a partitioning of the network. This definition is the one we follow in this work. (The definition of community can indeed vary depending on the field of study or the application.) As opposed to clustering methods which commonly involve a given number of clusters, communities within networks are usually unknown, can be of unequal size and density and often have hierarchies [10, 25]. Finding such partitioning can give information on the underlying structure of a network and its functioning. It can also be used as a more compact representation of the network, for instance for visualisations.

The detection of a community structure in networks can be divided in two problems. The first problem is algorithmic: "How to partition a graph?" The second problem is more semantic related: "How to measure the quality of a partition?" which requires an accepted definition of a community such as the one given above. These two problems can be seen as separate in some methods such as [30] where a fairly generic aggregating algorithm optimises a partition quality criterion. Other algorithms such as [4] still use the same criterion but exploit the topology of the network based on what the criterion is trying to achieve and attempt to guide the aggregation towards better solutions than a more generic aggregation method would produce.

The quality of a partition can be measured using several functions. To date the most common and explored measure is *modularity* [33]. Regarding the partitioning algorithm, partitioning graphs is an NP-hard task [10] and hence finding the optimum community partition is of similar complexity. Heuristics based algorithms have thus been devised to provide acceptable solutions at a reduced complexity. Considering the sizes of some real-world networks (e.g. web, protein networks) a lot of effort has been put into finding efficient algorithms able to deal with larger and larger networks.

However, returning to the definition of communities, it has been shown that networks often have several levels of organisation [43], leading to different partitions for each level. Therefore communities in networks can be present at various scales (or resolutions) and can have a hierarchical structure. This multi-scale aspect cannot be handled by modularity optimisation alone [11, 10]. To address this issue methods have been provided to adapt modularity optimisation to multi-scale (multi-resolution) analysis using a tuning parameter [39, 3]. Other methods such as [40] consider intra clusters connectivity and provide a tuning parameter to aim at communities of various sizes. Yet the search for a partition quality function that acknowledges the multi-resolution nature of networks with appropriate theoretical foundations has received less attention. Recently, *stability* [8] was introduced as a new quality measure for community partitions. Based on this measure [24] presented an investigation of the use of stability as an optimisation criterion for multi-scale analysis. In this work we expand upon these initial results and present a broader investigation of stability optimisation involving heuristics for speed gain and/or accuracy as well as an application to overlapping community detection.

The next section provides a literature review presenting contributions to community detection relevant to this work. Then the concept of stability optimisation is presented followed by the introduction of optimisation heuristics. Experiments are then performed in order to assess our method and its variation as well as to compare them with other relevant methods found in the literature. An application to the detection of overlapping communities is then presented and assessed. The paper is concluded by discussing the potential of our approach and its implications for the field and future work.

## 2   Background

While several community partition quality measures exists [26, 8, 10], the most commonly found in the literature is modularity [33]. Given a partition into $c$ communities let $e$ be the community matrix of size $c \times c$ where each $e_{ij}$ gives the fraction of links going from a community $i$ to a community $j$ and $a_i = \sum_j e_{ij}$ the fraction of links connected to $i$. (If the graph is undirected, each $e_{ij}$ not on the diagonal should be given half of the edges connecting communities $i$ and $j$ so that the number of edges connecting the communities is given by $e_{ij} + e_{ji}$ [33].) Modularity $Q_M$ is the sum of the difference between the fraction of links within a partition linking to this very partition minus the expected value of the fraction of links doing so if edges were randomly placed:

$$Q_M = \sum_{i=1}^{c}(e_{ii} - a_i^2) \qquad (1)$$

One advantage of modularity is to impose no constraint on the shape of communities as opposed for instance to the clique percolation method [35] that defines communities as adjacent k-cliques thus imposing that each node in a community is part of a k-clique.

Modularity was initially introduced to evaluate partitions. However its use has broadened from partition quality measure to optimisation function and modularity optimisation is now a very common approach to community detection [30, 5, 31, 4]. (Recent reviews and comparisons of community detection methods including modularity optimisation methods can be found in [10, 23].) Modularity optimisation methods commonly start with each node placed in a different community and then successively merge the communities that maximise modularity at each step. Modularity is thus locally optimised at each step based on the assumption that a local peak should indicate a particularly good partition. The first algorithm of this kind was *Newman's fast algorithm* [30]. Here, for each candidate partition the variation in modularity $\Delta Q_M$ that merging two communities $i$ and $j$ would yield is computed as

$$\Delta Q_{M_{ij}} = 2(e_{ij} - a_i a_j) \quad (2)$$

where $i$ and $j$ are the communities merged in the new candidate partition. Computing only $\Delta Q_M$ minimises the computations required to evaluate modularity and leads to the fast greedy algorithm given in Algorithm 1. This algorithm enables the incremental building of a hierarchy

---

**Algorithm 1** Sketch of a greedy algorithm for modularity optimisation.

1. Divide in as many clusters as there are nodes
2. Measure modularity variation $\Delta Q_M$ for each candidate partition where a pair of clusters are merged
3. Select the network with the highest $\Delta Q_M$
4. Go back to step 2

---

where each new partition is the local optima maximising $Q_M$ at each step. It was shown to provide good solutions with respect to the original Girvan-Newman algorithm that performs accurately but is computationally highly demanding and is thus not suitable for large networks [33]. (Note that accuracy refers in this context to a high modularity value. Other measures, such as stability, might rank partitions differently.) Since then other methods have been devised such as [5] optimising the former method, another approach based on the eigenvectors of matrices [31] and the Louvain method [4]. The latter has shown to outperform in speed previous greedy modularity optimisation methods by reducing the number of intermediate steps for building the hierarchy. The method works in two steps that are repeated until no further aggregation is possible. The first step aggregates nodes by moving them from their community to a neighbour community if this move results in an increase of modularity. Once no move can provide any increase the second step consists in creating the graph of the communities. Then the method is reapplied from step one to this new graph.

Other approaches have looked at introducing biases to alter the behaviour of the modularity optimisation towards communities of various sizes. In [6], the authors observed that in [30] large communities are favoured at the expense of smaller ones biasing the partitioning towards a structure with a few large clusters which may not be an accurate representation of the network. They provided a normalised measure of $\Delta Q_M$ defined as

$$\Delta Q'_{M_{ij}} = max\left(\frac{\Delta Q_{M_{ij}}}{a_i}, \frac{\Delta Q_{M_{ij}}}{a_j}\right) \quad (3)$$

which aims at treating communities of different sizes equally.

These methods all rely on modularity optimisation. Yet modularity optimisation suffers from several issues. One issue is known as the resolution limit meaning that modularity optimisation methods can fail to detect small communities or over-partition networks [11] thus missing the most natural partitioning of the network. Another issue is that the modularity landscape admits a large number of structurally different high-modularity value solutions and lacks a clear global maximum value [14]. It has also been shown that random-graphs can have a high modularity value [15].

However not all methods rely on modularity. A popular non modularity-related method is known as InfoMap [41]. InfoMap considers information flow probabilities of random walks within a network to compress the network. The compression principle decomposes the network into modules that provide a coarse view of the graph. This view provides a first level of encoding. Then a second level of encoding is performed locally in each module. The content of each modules, namely a neighbourhood of nodes, provides the fine-grain view of the graph. This technique of compression is therefore based on a two level representation of the graph. The module level provides a partition of the network that can be interpreted in terms of communities as at highlights the structure of the network.

So far the methods presented only return one best partition for a given network. Yet it is commonly acknowledged that networks often have several levels of organisation [43]. Therefore community detection methods should be able to identify different partitions at various scales. In this respect several methods have been proposed. In [39], modularity optimisation is modified by using a scalar parameter $\gamma$ in front of the null term (the fraction of edges connecting vertices of a same community in a random graph) turning equation (1) into

$$Q_{M_\gamma} = \sum_i (e_{ii} - \gamma a_i^2) \qquad (4)$$

where $\gamma$ can be varied to alter the importance given to the null term (modularity optimisation is found for $\gamma = 1$). In [3], modularity optimisation is performed on a network where each node's strength has been reinforced with self loops. Considering the adjacency matrix $A$, modularity optimisation is performed on $A + rI$ where $I$ is the identity matrix and $r$ is a scalar:

$$Q_{M_r} = Q_M(A + rI) \qquad (5)$$

Varying the value of $r$ enables the detection of communities at various coarseness levels (modularity optimisation is found for $r = 0$). With their resolution parameter, the two latter methods enable a multi-scale network analysis.

Another multi-scale method, not relying on modularity, was introduced in [40]. The model uses no null factor and is therefore not subject to the resolution limit found in modularity. The quality of a partition is expressed as follows:

$$Q_H(\gamma) = -\frac{1}{2} \sum_{i \neq j} (A_{ij} - \gamma J_{ij}) \qquad (6)$$

where $A$ is the adjacency matrix and $J_{ij} = 1 - A_{ij}$ (i.e. $J$ is the complement of the adjacency matrix $A$) and $\gamma$ the resolution parameter. The models therefore considers the amount of connections within clusters less the connections missing to get fully connected clusters. $\gamma$ varies the importance of the missing connections. A small $\gamma$ value will favour large clusters while a large $\gamma$ value will favour dense clusters.

Recently, a new partition quality measure called *stability* was introduced in [8]. The stability of a graph considers the graph as a Markov chain where each node represents a state and each edge a possible state transition. Let $n$ be the number of nodes, $m$ the number of edges, $A$ the $n \times n$ adjacency matrix containing the weights of all edges (the graph can be weighted or not), $d$ a size $n$ vector giving for each node its degree (or strength for a weighted network) and $D = diag(d)$ the corresponding diagonal matrix. The stability of a graph considers the graph as a Markov chain where each node represents a state and each edge a possible state transition. The chain distribution is given by the stationary distribution

$$\pi = \frac{d}{2m} \qquad (7)$$

Also let $\Pi$ be the corresponding diagonal matrix $\Pi = diag(\pi)$. The transition between states is given by the $n \times n$ stochastic matrix

$$M = D^{-1}A \qquad (8)$$

Assuming a community partition, let $H$ be the indicator matrix of size $n \times c$ giving for each node its community. The clustered auto-covariance matrix at Markov time $t$ is defined as:

$$R_t = H^T(\Pi M^t - \pi^T \pi)H \qquad (9)$$

Stability at time $t$ noted $Q_{S_t}$ is given by the trace of $R_t$ and the global stability measure $Q_S$ considers the minimum value of the $Q_{S_t}$ over time from time $0$ to a given upper bound $\tau$:

$$Q_S = \min_{0 \leq t \leq \tau} trace(R_t) \qquad (10)$$

This model can be extended to deal with real values of $t$ by using the linear interpolation:

$$R_t = (c(t) - t) \cdot R(f(t)) + (t - f(t)) \cdot R(c(t)) \qquad (11)$$

where $c(t)$ returns the smallest integer greater than $t$ and $f(t)$ returns the greatest integer smaller than $t$. This is useful to investigate for instance time values between 0 and 1 in which case the equation becomes simply:

$$R_t = (1 - t) \cdot R(0) + t \cdot R(1) \qquad (12)$$

It was indeed shown in [8] that the use of Markov time with values between 0 and 1 enables detecting finer partitions than those detected at time 1 and above.

Also, this model can be turned into a continuous time Markov process by using the expression $e^{(M-I)t}$ in place of $M^t$ (where $e$ is the exponential function) [8]. Several Markov processes could be considered here as discussed in [22].

Stability has been introduced as a measure to evaluate the quality of a partition hierarchy and has been used to assess the results of various modularity optimisation algorithms. Further mathematical foundations have been presented in [22, 21]. Stability has been shown to unify some known clustering heuristics including modularity. Based on this theoretical work, [24] presented a stability optimisation method similar to the fast modularity optimisation method from [30]. The work showed that stability can be optimised similarly to modularity optimisation.

While related approaches such as [3, 39] also offer a multi-scale analysis with their respective parameters, these methods offer a tuneable version of modularity optimisation by modifying the importance of the null factor or by adding self-loops to nodes. Such analysis remains based on a one step random walk analysis of the network with modifications of its structure. [40] also considers a very strict topology criterion by aiming at fully connected communities. In contrast, stability optimisation enables random walks of variable length defined by the Markov time thus exploiting thoroughly the actual topology of the network similarly to an information flow through a network. As communities reflect the organisation of a network, and hence its connectivity, this approach seems to be more suitable.

The next section presents the stability optimisation method. The paper then develops further the idea by presenting several optimisation heuristics that enable significant gains in speed and potentially in accuracy. Then the method is applied to the detection of overlapping communities. All the methods are tested against several datasets and compared with one another as well as with other popular methods.

## 3 Stability Optimisation

### 3.1 Principle

As discussed in [8], studying the stability of a partition along the Markov time can help addressing the partition scale issue and the optimal community identification. The results from the authors indeed show with the stability curve that the clustering varies depending on the time window during which the Markov time is considered. From there, [24] used the Markov time as a resolution parameter in a greedy optimisation context where stability is used as the optimisation criterion. The principle is the following.

Let $A_t = DM^t$, considering equation (9), the autocovariance can also be expressed as

$$R_t = H^T(\Pi M^t - \pi^T \pi)H = H^T(\frac{1}{2m}A_t - \pi^T \pi)H = \frac{1}{2m}H^T A_t H - H^t \pi^T \pi H \qquad (13)$$

The trace of the autocovariance can then be expressed as the modularity of the graph with adjacency matrix $A_t$. The community matrix for $A_t$ is then noted $e_t$.

$$trace(R_t) = trace(\frac{1}{2m}H^T A_t H - H^t \pi^T \pi H) = \frac{1}{2m}trace(H^T A_t H) - trace(H^t \pi^T \pi H) \qquad (14)$$

Using this expression, this trace can then be expressed using two forms commonly found in the modularity literature. The first one uses the adjacency matrix with nodes indices $i$ and $j$ as well as the $\delta(i,j)$ function returning 1 if $i$ and $j$ belong to the same community, zero otherwise:

$$trace(R_t) = \frac{1}{2m}\sum_{i,j} A_{t_{ij}}\delta(i,j) - (\frac{1}{2m})^2 \sum_{i,j} d_i d_j \delta(i,j) = \frac{1}{2m}\sum_{i,j}(A_{t_{ij}} - \frac{d_i d_j}{2m})\delta(i,j) \qquad (15)$$

The second form uses the previously defined community matrix, noted here $e_t$ for time $t$:

$$trace(R_t) = \sum_{i=1}^{c} e_{t_{ii}} - \sum_{i=1}^{c} a_i^2 = \sum_{i=1}^{c}(e_{t_{ii}} - a_i^2) \qquad (16)$$

In this work we will use the latter expression. Stability at time $t$ is therefore defined as:

$$Q_{S_t} = trace(R_t) = \sum_{i=1}^{c}(e_{t_{ii}} - a_i^2) \qquad (17)$$

This is the analogue of equation (1) for $A_t$. Therefore the optimisation of stability at time $t$ only is equivalent to the optimisation of modularity taking the adjacency matrix $A_t$. These results are for the Markov chain model. The continuous time model would use the adjacency matrix $A_t = De^{(M-I)t}$. Considering stability at time $t$ as the modularity of a graph given by the adjacency matrix $A_t$ allows the modularity optimisation techniques to be applied to stability.

Stability optimisation then becomes a broader measure where modularity is the special case $t = 1$ in the Markov chain.

Greedy modularity optimisation is based on computing the change in modularity between an initial partition and a new partition where two clusters have been merged. The change in modularity when merging communities $i$ and $j$ is given by equation (2) and similarly the change in stability at time $t$ is

$$\Delta Q_{S_{t_{ij}}} = 2(e_{t_{ij}} - a_i a_j) \tag{18}$$

Following equation (10) the new $Q_S$ candidate value $Q'_S$ is:

$$Q'_S = \min_{0 \leq t \leq \tau}(Q_{S_t} + \Delta Q_{S_t}) \tag{19}$$

Other modularity optimisation methods may use different ways to compute $\Delta Q_M$ based on the way they aggregate communities (e.g. Louvain method from [4]). Considering that stability optimisation at time t can be seen as modularity optimisation for $A_t$, these expressions would also be valid for computing $\Delta Q_S$.

## 3.2 Greedy Optimisation

Following the steps from Algorithm 1, a similar stability optimisation method can be derived [24]. At each clustering step, the partition with the best $Q'_S$ value is kept and $Q_S$ is then updated as $Q_S = Q'_S$. For computational reasons the time needs to be sampled between 0 and $\tau$. Markov time can be sampled linearly or following a log scale. The latter is usually more efficient for large time intervals.

All the matrices $e_t$ are computed in the initialisation step of the algorithm and then updated by successively merging the lines and columns corresponding to the communities merged together. This leads to the greedy stability optimisation (GSO) algorithm given in Algorithm 2 from [24], based on the principle of Algorithm 1. Depending on the boundaries considered for the Markov time the partition returned by the algorithm will vary. Indeed, the larger the Markov time window, the longer in time a partition must keep a high stability value to get a high overall stability value, as defined in equation (10). The Markov time thus acts as a resolution parameter.

Compared to Newman's fast algorithm the additional cost of stability computation and memory requirement is proportional to the number of times considered in the Markov time window. For each time $t$ considered in the computation, a matrix $e_t$ must be computed and kept in memory. Let $n$ be the number of nodes in a network, $m$ the number of edges and $s$ the number of time steps required for stability computation. The number of merge operations needed to terminate are $n - 1$. Each merging operation requires to iterate through all edges, hence $m$ times, and for each edge to compute the stability variation $s$ times. The computation of each $\Delta Q$ can be performed in constant time. In a non optimised implementation the merging of communities $i$ and $j$ can be performed in $n$ steps, hence the complexity of the algorithm would be $\mathcal{O}(n(m.s + n))$. However the merging of communities $i$ and $j$ really consists in adding the edges of community $j$ to $i$ and then delete $j$. To do so there is one operation per edge. The size of a cluster at iteration $i$ is given by:

$$cs(i) = \frac{n}{n - i} \tag{20}$$

and therefore the average cluster size over all iterations is

$$\bar{cs} = \frac{1}{n} \cdot \sum_{i=1}^{n} cs(i) = \sum_{i=1}^{n} \frac{1}{i} \approx ln(n) + \gamma \tag{21}$$

**Algorithm 2** Greedy Stability Optimisation (GSO) algorithm taking in input an adjacency matrix and a window of Markov times, and returning a partition and its stability value.

>   Divide in as many communities as there are nodes
>   Set this partition as current partition $C_{cur}$ and as best known partition $C$
>   Set its stability value as best known stability $Q$
>   Set its stability vector (stability values at each Markov time) as current stability vector $QV$
>   Compute initial community matrix $e$
>   Compute initial community matrices at Markov times $e_t$
>   **while** 2 communities at least are left in current partition: $length(e) > 1$ **do**
>       Initialise best loop stability $Q_{loop} \leftarrow -\infty$
>       **for all** pair of communities with edges linking them: $e_{ij} > 0$ **do**
>           **for all** times $t$ in time window **do**
>               Compute $dQV(t) \leftarrow \Delta Q_{S_t}$
>           **end for**
>           Compute partition stability vector: $QV_{tmp} \leftarrow QV + dQV$
>           Compute partition stability value by taking its minimum value: $Q_{tmp} \leftarrow min(QV_{tmp})$
>           **if** current stability is the best of the loop: $Q_{tmp} > Q_{loop}$ **then**
>               $Q_{loop} \leftarrow Q_{tmp}$
>               $QV_{loop} \leftarrow QV_{tmp}$
>               Keep in memory best pair of communities $(i, j)$
>           **end if**
>       **end for**
>       Compute $C_{cur}$ by merging the communities $i$ and $j$
>       Update matrices $e$ and $e_t$ by merging rows $i$ and $j$ and columns $i$ and $j$
>       Set current stability vector to best loop stability vector: $QV \leftarrow QV_{loop}$
>       **if** best loop stability higher than best known stability: $Q_{loop} > Q$ **then**
>           $Q \leftarrow Q_{loop}$
>           $C \leftarrow C_{cur}$
>       **end if**
>   **end while**
>   **return** best found partition $C$ and its stability $Q$

where $\gamma$ is the Euler-Mascheroni constant. As the number of edges is bounded by the number of nodes squared, the algorithm can be implemented with the complexity $\mathcal{O}(n(m.s + ln^2(n)))$ provided the appropriate data structures are used to exploit this, as discussed in [5]. As $s$ should be a low value, the complexity is $\mathcal{O}(n(m + ln^2(n)))$. It should however be noted that the technique used to compute the matrices $M^t$ can bring additional complexity, whether using a naive approach, a fast matrix multiplication algorithm such as Strassen's [44], other data structures and/or approximation techniques. The optimisation of this task will not be addressed in this work.

## 3.3 Analysis

Frequently in the literature modularity is used to rank the quality of algorithms. However as this work optimises stability, using modularity is inappropriate. Modularity has also several drawbacks as previously mentioned (e.g. random-graphs can have high modularity [15], modularity has a resolution limit [11]). Finally as our method enables multi-scale network analysis and thus considers several relevant partitions per network, modularity is ill-suited.

In the absence of any knowledge of a network, the analyst would look for partitions that are consistently found on some intervals of the Markov time. These will be called *stable* partitions. First, such partitions should have the same number of communities. (In some cases one may look for very similar partitions and not identical ones though.) However, while successive partitions for successive Markov times are likely to have some similarity, the number of communities does not necessarily reflect the actual partitioning of a network. Therefore an additional measure that considers the actual information contained in the partitions is required in order to better identify stable partitions. This can be achieved by using the *normalised mutual information* (NMI) [12] which has proved to be reliable [7]. The NMI between two partitions $A$ and $B$ is defined as:

$$NMI(A,B) = \frac{-2\sum_{i=1}^{c_A}\sum_{j=1}^{c_B} C_{ij} log(\frac{C_{ij} \cdot n}{C_{i*} C_{*j}})}{\sum_{i=1}^{c_A} C_{i*} log(\frac{C_{i*}}{n}) + \sum_{j=1}^{c_B} C_{*j} log(\frac{C_{*j}}{n})}$$

where $n$ is the number of nodes, $c_A$ is the number of communities in $A$, $c_B$ is the number of communities in $B$, $C$ is the confusion matrix where $C_{ij}$ is the number of nodes of community $i$ in $A$ that are in community $j$ in $B$, $C_{i*}$ is the sum of elements in row $i$, $C_{*j}$ is the sum of elements in column $j$. If $A = B$ then $NMI(A,B) = 1$. On the contrary if $A$ and $B$ are completely different (including if $A$ or $B$ classifies everything in one single cluster) then $NMI(A,B) = 0$.

Computing the NMI between successive partitions (i.e. one at time $t$ and the next one at time $t + dt$) provides additional information beyond the number of communities found in each partition and enables a fine detection of stable partitions.

## 4 Heuristics

### 4.1 Time-window Optimisation

As investigated in [24] a large time window does not imply many intermediate time values. While the full mathematical definition of stability considers all Markov times in a given interval, all Markov times may not be crucial to a good (or even exact) approximation of stability. The fastest way to approximate stability is to compute it with only one Markov time value. As stability tends to decrease as the Markov time increases, we are seeking whether the following approximation can be made:

$$Q_S = \min_{0 \leq t \leq \tau} trace(R_t) \approx trace(R_\tau) \qquad (22)$$

The need for considering consecutive time values in the computation of stability addresses an issue encountered within random walks. Considering for instance a graph with three nodes $a$, $b$ and $c$ with an edge between nodes $a$ and $b$ and between nodes $b$ and $c$. Using a Markov time of 2 only (i.e. a random walk of 2 steps with no consideration of the first step) starting from $a$ there would be no transition between $a$ and $b$ as after one step from $a$ the random walker would be in $b$ and then it could only go back to $a$ or walk to $c$. However, the more densely connected the clusters, the less likely this situation is to happen as many paths can be taken to reach each node. The work from [24] showed that optimising stability based on equation (22) provides results very similar in accuracy to those found by optimising the full stability (i.e. using a finely sampled time-window) while providing a significant gain in speed.

In order to exploit these results algorithmically, the greedy stabilisation at one Markov time becomes very similar to a modularity optimisation algorithm. The input time window becomes a single time value as opposed to an array of time values and the computation of the stability can be simplified, as given in Algorithm 3.

### 4.2 Randomisation

Another approach for speed optimisation is to randomise the aggregation algorithm. Considering Algorithm 2, looking for the best pair in all the possible pairs in the network is time consuming. The merging of two communities affects the stability based only on the local structure affected

**Algorithm 3** Modification of the Greedy Stabilisation Algorithm algorithm from Algorithm 2 optimised for using a single Markov time value.

---
Divide in as many communities as there are nodes
Set this partition as current partition $C_{cur}$ and as best known partition $C$
Set its stability value as best known stability $Q$ and current partition stability $Q_{cur}$
Compute initial community matrices $e$ and $e_t$
**while** 2 communities at least are left in current partition: $length(e) > 1$ **do**
    Initialise best loop stability variation $\Delta Q_{loop} \leftarrow -\infty$
    **for all** pair of communities with edges linking them: $e_{ij} > 0$ **do**
        Compute local variation of stability $\Delta Q_{tmp}$
        **if** current stability variation is the best of the loop: $\Delta Q_{tmp} > \Delta Q_{loop}$ **then**
            $\Delta Q_{loop} = \Delta Q_{tmp}$
            Keep in memory best pair of communities $(i, j)$
        **end if**
    **end for**
    Compute $C_{cur}$ by merging the communities $i$ and $j$
    Update matrices $e$ and $e_t$ by merging rows $i$ and $j$ and columns $i$ and $j$
    Add to current stability the best loop stability variation: $Q_{cur} \leftarrow Q_{cur} + \Delta Q_{loop}$
    **if** current stability higher than best known stability: $Q_{cur} > Q$ **then**
        $Q \leftarrow Q_{cur}$
        $C \leftarrow C_{cur}$
    **end if**
**end while**
**return** best found partition $C$ and its stability $Q$

---

by this change. Therefore looking at all the possible pairs to merge in the entire network at each pass may not always be necessary and yet is significantly time consuming. The most important at each pass is rather to select the best pair within a set of pairs affecting the same area in a network. This idea has been developed in [34] for modularity optimisation. The same principle can be adapted to stability optimisation, as given in Algorithm 4 below. At each pass the algorithm selects $k$ communities, with $k$ a parameter to be set, and merges the best pair found within these $k$ communities instead of within the whole network. We set here $k$ using the best values found in [34]: $k = 1$ during the first half of the community merging process in Algorithm 2 and then taking $k = 2$ for the second half. This approach significantly reduces the time required for each pass. Before the complexity of a pass was $\mathcal{O}(m)$ as each pass was iterating over all edges. With this heuristic each pass takes $k$ communities and checks the edges linking it to other communities. As found earlier, the average cluster size is in $\mathcal{O}(ln(n))$ and the number of edges is thus bound by $\mathcal{O}(ln^2(n))$. As a result the complexity of a pass is in $\mathcal{O}(k \cdot ln^2(n)) = \mathcal{O}(ln^2(n))$. Therefore the algorithm complexity becomes $\mathcal{O}(n \cdot ln^2(n))$.

Another advantage of exploring randomly only a subpart of the graph at each pass is that it enables the formation of clusters concurrently. By considering all pairs throughout the network like in Algorithm 1 or Algorithm 2 there is a risk of formation of large clusters that may absorb vertices rather than allowing the formation of new communities.

### 4.3 Multi-step Aggregation

In [42] the authors suggest a different way to avoid the formation of large clusters by using a multi-step approach that merges several pairs of candidate clusters per pass, thus greatly promoting the concurrent formation of clusters. The method has also the advantage of reducing the number of passes from $n - 1$ to $\frac{n-1}{k}$ with $k$ the number of cluster merging per pass. The same principle can be applied for stability optimisation, as given in Algorithm 5. The complexity is therefore similar but with a $k$ division factor which with the time optimised version is $\mathcal{O}(\frac{n}{k}(m + ln^2(n)))$. While in the same order of complexity as the non-optimised version the $k$ factor enables a gain

**Algorithm 4** Randomised Greedy Stability Optimisation (RGSO) algorithm taking in input an adjacency matrix and a window of Markov times, and returning a partition and its stability value.

    Divide in as many communities as there are nodes
    Set this partition as current partition $C_{cur}$ and as best known partition $C$
    Set its stability value as best known stability $Q$
    Set its stability vector (stability values at each Markov time) as current stability vector $QV$
    Compute initial community matrix $e$
    Compute initial community matrices at Markov times $e_t$
    **while** 2 communities at least are left in current partition: $length(e) > 1$ **do**
        Initialise best loop stability $Q_{loop} \leftarrow -\infty$
        **if** $nb\_lines(e) < nb\_lines(adj)/2$ **then**
            k = 2
        **else**
            k = 1
        **end if**
        **for** $c = 1$ to $k$ **do**
            Select random community i (different from previously picked if $k > 1$)
            **for all** communities $j$ sharing edges with $i$ **do**
                **for all** times $t$ in time window **do**
                      Compute $dQV(t) \leftarrow \Delta Q_{S_t}$
                **end for**
                Compute partition stability vector: $QV_{tmp} \leftarrow QV + dQV$
                Compute partition stability value by taking its minimum value: $Q_{tmp} \leftarrow min(QV_{tmp})$
                **if** current stability is the best of the loop: $Q_{tmp} > Q_{loop}$ **then**
                    $Q_{loop} \leftarrow Q_{tmp}$
                    $QV_{loop} \leftarrow QV_{tmp}$
                    Keep in memory best pair of communities $(i, j)$
                **end if**
            **end for**
        **end for**
        Compute $C_{cur}$ by merging the communities $i$ and $j$
        Update matrices $e$ and $e_t$ by merging rows $i$ and $j$ and columns $i$ and $j$
        Set current stability vector to best loop stability vector: $QV \leftarrow QV_{loop}$
        **if** best loop stability higher than best known stability: $Q_{loop} > Q$ **then**
            $Q \leftarrow Q_{loop}$
            $C \leftarrow C_{cur}$
        **end if**
    **end while**
    **return** best found partition $C$ and its stability $Q$

in overall execution time that can be significant.

One drawback of this method when optimising modularity is that one cannot know in advance the best value for the parameter $k$ [42]. However as we are using stability and not modularity, we are mainly interested in the behaviour when reaching stable partitions. One insight is that, compared to community detection using modularity, the detection of stable partitions may be less sensitive to the tuning of $k$ as these partitions are persistent through a window of time and therefore already show a form of robustness to parameter setup. This is investigated in the following series of experiments.

## 4.4 Time-window Optimisation combined with the Louvain Method

The previous optimisation methods explored a variation of the greedy stability optimisation where only one Markov time is considered instead of a time window. In addition it has been shown that optimising stability for time $t$ is equivalent to optimising modularity of the graph with adjacency matrix $A_t$, as given in equation (17). Therefore using Newman's greedy modularity

**Algorithm 5** Multi-Step Greedy Stability Optimisation (MSGSO) algorithm taking in input an adjacency matrix, a window of Markov times and a number of pairs $k$ to merge per iteration, and returning a partition and its stability value.

---
Divide in as many communities as there are nodes
Set this partition as current partition $C_{cur}$ and as best known partition $C$
Set its stability value as best known stability $Q$
Set its stability vector (stability values at each Markov time) as current stability vector $QV$
Compute initial community matrix $e$
Compute initial community matrices at Markov times $e_t$
**while** 2 communities at least are left in current partition: $length(e) > 1$ **do**
    Set the pair list to an empty list
    **for all** pair of communities with edges linking them: $e_{ij} > 0$ **do**
        **for all** times $t$ in time window **do**
            Compute $dQV(t) \leftarrow \Delta Q_{S_t}$
        **end for**
        Compute partition stability vector: $QV_{tmp} \leftarrow QV + dQV$
        Compute partition stability value by taking its minimum value: $Q_{tmp} \leftarrow min(QV_{tmp})$
        **if** current stability is the best of the loop: $Q_{tmp} > Q_{loop}$ **then**
            $Q_{loop} \leftarrow Q_{tmp}$
            $QV_{loop} \leftarrow QV_{tmp}$
            Add the pair $(i, j)$ with the values $Q_{tmp}$ and $dQV$ to the pair list
        **end if**
    **end for**
    Sort the pair list by their stability values $Q_{tmp}$
    **for** $k$ times (the number of pairs to merge per iteration) **do**
        Select best pair $(i, j)$ in the pair list
        Compute $C_{cur}$ by merging the communities $i$ and $j$
        Update matrices $e$ and $e_t$ by merging rows $i$ and $j$ and columns $i$ and $j$
        Set current stability vector to best loop stability vector: $QV \leftarrow QV_{loop}$
        **if** best loop stability higher than best known stability: $Q_{loop} > Q$ **then**
            $Q \leftarrow Q_{loop}$
            $C \leftarrow C_{cur}$
        **end if**
        Remove from the pair list all pairs that share a node with the selected pair
    **end for**
**end while**
**return** best found partition $C$ and its stability $Q$

---

optimisation would be equivalent to using Algorithm 2. Yet, instead of Newman's algorithm any other modularity optimisation method can potentially be used, such as the Louvain method [4] previously mentioned and interesting for its execution speed. The Markov time thus remains the resolution parameter to compute the matrix $A_t$ but enables the Louvain method to process the resulting network without modifying its code. This offers an alternative algorithm to optimise stability. Comparing the results of this combination with the other methods to optimise stability can also enable to evaluate how robust are the detected stable partitions with respect to the aggregation algorithm in addition to the Markov time.

## 5 Experiments

This section presents sets of experiments that were run to assess the methods presented in this paper. The community detection algorithms were implemented in Matlab[1] except for the code of the Louvain method downloaded from the authors website[2] (we used their hybrid C++ Matlab implementation). All experiments were run using Matlab R2011a under MacOS X on an iMac 3.06GHz Intel Core i3.

---
[1] The code developed for this work is available on request. More can also be found at `http://www.elemartelot.org`.

## 5.1 Test Networks

The networks considered for the experiments are two synthetic and four real-world data networks[3] that have been used as benchmarks in the literature to assess community detection algorithms. The networks have been chosen for their respective properties (e.g. multi-scale, scale-free) and popularity that enable an assessment of our method and comparisons with other approaches. While selecting very large networks can demonstrate speed efficiency, the results are commonly ranked using modularity which as previously discussed is not suitable for this work. We therefore deemed appropriate to use here networks of smaller size with some knowledge of their structure or content used for the evaluation of the results, similarly to what has been done in related work [39, 3, 4, 40].

The following two synthetic datasets are used:

**Ravasz and Barabási's scale-free hierarchical network:** This network was presented in [38] and defines a hierarchical network of 125 nodes as shown in Figure 1(a). It is hereafter referred to as *RB-125*. The network is built iteratively from a small cluster of 5 densely linked nodes. A node at the centre of a square is connected to 4 others at the corners of the square themselves also connected to their neighbours. Then 4 replicas of this cluster are generated and placed around the first one, producing a 25 nodes network. The centre of the central cluster is linked to the corner nodes of the other clusters. This process is repeated again to produce the 125 nodes network. The structure can be seen as 25 clusters of 5 nodes or 5 clusters of 25 nodes.

**Arenas et al's homogeneous in degree network:** This network taken from [2] and named H13-4 is a two hierarchy levels network of 256 nodes organised as follow. 13 edges are distributed between the nodes of each of the 16 communities of the first level (internal community) formed of 16 nodes each. 4 edges are distributed between nodes of each of the 4 communities of the second level (external community) formed of 64 nodes each. 1 edge per node links it with any community of the rest of the network. The network is presented in Figure 1(b).

The following real-data networks are used:

**Zachary's karate club:** This network is a social network of friendships between 34 members of a karate club at a US university in 1970 [45]. Following a dispute the network was divided into 2 groups between the club's administrator and the club's instructor. The dispute ended in the instructor creating his own club and taking about half of the initial club with him. The network considered here is the unweighted version of the network, and is undirected. The network can hence be divided into 2 main communities, as shown in Figure 2(a). A division into 4 communities has also been acknowledged [29].

**Lusseau et al's dolphins social network:** This network is an undirected social network resulting from observations of a community of 62 bottle-nose dolphins over a period of 7 years [28]. Nodes represent dolphins and edges represent frequent associations between dolphin pairs occurring more often than expected by chance. Analysis of the data revealed 2 main groups as shown in Figure 2(b) and a further division can be made into 4 groups [27].

**American college football dataset:** This dataset contains the network of American football games between Division I colleges during regular season Fall 2000 [13]. The 115 nodes represent teams and the edges represent games between 2 teams. The teams are divided into 12 groups containing around 8-12 teams each and games are more frequent between members of the same group (teams play on average seven intragroup games and four intergroup games). Also teams that are geographically close but belong to different groups are more likely to play one another than teams separated by a large distance. Therefore in this dataset the groups can be considered as known communities as their nodes should be more interconnected than with the rest of the network.

---
[2]http://sites.google.com/site/findcommunities/
[3]Available from http://www-personal.umich.edu/~mejn/netdata/

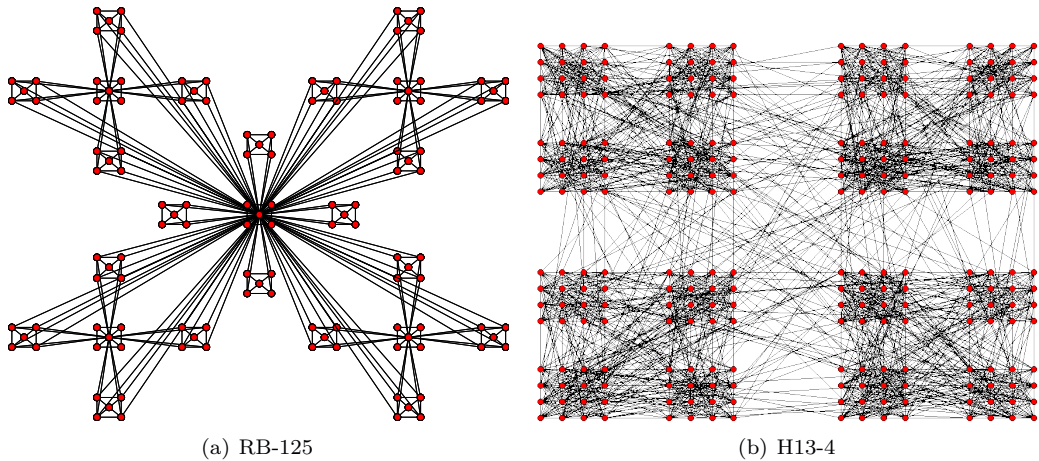

(a) RB-125    (b) H13-4

Figure 1: (a) Hierarchical scale free network generated in 3 steps producing 125 nodes at step 3 [38]. (b) Network presented in [2] made of 256 nodes organised in two hierarchical levels with 16 communities of 16 nodes for the first level and 4 communities of 64 nodes for the second level.

**Les Misérables:** This dataset taken from [19] represents the co-appearance of 77 characters in Victor Hugo's novel *Les Misérables*. Two nodes share an edge if the corresponding characters appear in a same chapter of the book. The values on the edges are the number of such co-appearances.

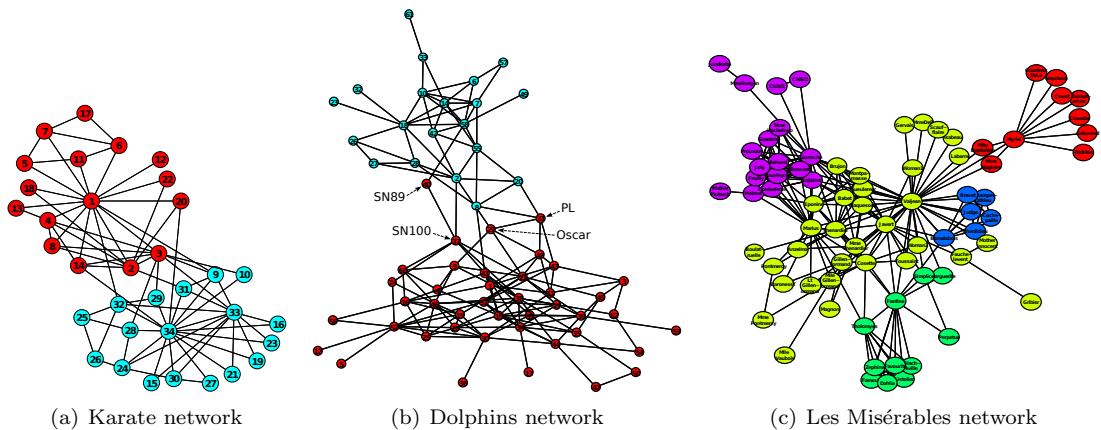

(a) Karate network    (b) Dolphins network    (c) Les Misérables network

Figure 2: Two real-world data networks: (a) Zachary's karate club network [45], (b) Lusseau's dolphins network [28], (c) Characters co-appearance network taken from the novel Les Misérables. The colours represent different identified communities.

## 5.2 Assessment and Comparison with Other Approaches

The first set of experiments assesses the greedy stability optimisation method without heuristic (as presented in Section 3). The method is tested on the one hand against various datasets and on the other hand against other relevant approaches. The methods used for comparison are Newman's fast algorithm [30], Danon et al's method [6], the Louvain method [4], Infomap [41], Reichardt and Bornholdt's method [39], Arenas et al's method [3] and Ronhovde and Nussinov's method [40]. Both Reichardt and Bornholdt's, Arenas et al's and Ronhovde and Nussinov's methods use a resolution parameter that enables multi-scale community detection based on this parameter. The Louvain method also returns a succession of partitions (not tuneable). The other methods are not multi-scale and hence return only one partition.

First to assess and compare the algorithms used in this work we use the established knowledge of the communities for each network to assess the algorithms results. We then use the normalised mutual information to analyse in more detail the stability optimisation results.

### 5.2.1 Comparative Results

The networks are analysed by the seven aforementioned community detection methods (4 non multi-scale, 3 multi-scale) and our stability optimisation algorithm for which both discrete and continuous Markov time versions are used.

Table 1 provides the results of the community detection methods that have no resolution parameter. The results show that these methods do not necessarily find the most meaningful partitions and that different methods can identify different partitions. Yet an unexpected result is not necessarily meaningless. For example in the RB-125 network, the central node of the central community has many more connections than the other nodes and more connections outside its community than inside. Therefore this node can be seen as a community on its own. While it is not the initially expected answer, this partition is relevant and reflects the situation of the node according to some optimisation criteria.

Table 1: Number of detected communities by Newman's fast algorithm, Danon et al's, Louvain and InfoMap methods on the presented networks. The identified division(s) known for these networks are also indicated ('-' indicates that there is no clear a priori knowledge). The Louvain method returns a hierarchy of partitions, given in order.

| Algorithm | RB-125 | H13-4 | Karate | Dolphins | Football | Miserables |
|---|---|---|---|---|---|---|
| Fast Newman | 6 | 4 | 3 | 4 | 6 | 5 |
| Danon | 6 | 4 | 4 | 4 | 6 | 6 |
| Louvain | 30, 10, 6 | 12, 4 | 6, 4 | 10, 5 | 12, 10 | 9, 6 |
| InfoMap | 22 | 13 | 3 | 6 | 12 | 10 |
| Identified | 5, 25 | 4, 16 | 2, 4 | 2, 4 | 12 | - |

Figure 3 plots the results of the tuneable methods along their respective parameter values. For Reichardt and Bornholdt's as well as Ronhovde and Nussinov's algorithm, the x-axis represents $10\gamma$. For Arenas et al's the x-axis represents $r - r_0$ where $r_0 = -\frac{m}{n}$ with $m$ the number of edges and $n$ the number of nodes. (See [3] for details. The authors use the lower bound $r_{asymp} = -\frac{2m}{n}$ but we found experimentally that values of $r$ below $r_0$ were irrelevant.) The value of $r_0$ is calculated for each network. For our algorithm, the x-axis represents the Markov time $t$. The time window sampling is done from time 0 to 100 with a step of 0.05 between within $[0, 2]$, a step of 0.25 within [2, 10] and a step of 1 afterwards. The steps between successive values of the parameter are 0.05 for Arenas and $\frac{0.05}{10} = 0.005$ for Reichardt and Bornholdt's and Ronhovde and Nussinov's methods (as the x-axis represents $10\gamma$).

(a) RB-125 network

(b) H13-4 network

(c) Karate club network

(d) Dolphins social network

(e) American football network

(f) Les Misérables characters network

Figure 3: Number of partitions returned by Reichardt and Bornholdt's, Arenas et al's, Ronhovde and Nussinov's and our stability optimisation methods. The x-axis represents $10\gamma$ for Reichardt and Bornholdt's and Ronhovde and Nussinov's methods, $r - r_0$ for Arenas et al's and $t$ for ours. The dotted horizontal lines indicate known partitions size: (a) 5 and 25, (b) 4 and 16, (c) 2 and 4, (d) 2 and 4, (e) 12.

Considering only stability optimisation we can observe that the two Markov processes behave in a very similar way. For this reason we will only comment on the discrete time version, the same being true for the other model. From Figure 3 it can be observed that the behaviour of our method is opposite to those of Reichardt and Bornholdt's and Arenas et al's methods and also very different from Ronhovde and Nussinov's method. At low values of $t$ partitions are small (at $t = 0$ our method finds as many partitions as there are nodes). Then as $t$ increases, partitions tend to become larger. Conversely, the three other methods start by partitioning in large clusters when their parameter is minimal. Then they find finer partitions as their parameter increases. However it is noteworthy that the progression of Ronhovde and Nussinov's method can be different from the two others, for instance in Figure 4(d) or Figure 3(f).

Considering Figure 3(a) for instance, after $t = 4$, the result of our algorithm remains stable on 5 clusters that represent the most *stable* partition. A partition in 6 elements can also be found and corresponds to the 5 large communities with the centre of the central community in a separate community. Also, when $t \in [0.1, 0.2]$ the algorithm finds a partition in 26 communities (25 small communities plus central node on its own). Arenas et al's method detects the 5 communities around about $r - r_0 \in [1.5, 2.5]$. Then it grows to stabilise on 25 communities around $r - r_0 \in [9, 25]$ and then goes up to 26 communities and more. Reichardt and Bornholdt's method stabilises around 5 communities around $\gamma = [0.5, 0.9]$ and around 26 communities around $\gamma = 3.6$ and onwards. (Note that it may still go higher for greater values of $\gamma$.) Ronhovde and Nussinov's method mainly stabilises on a partition of size 25 but also detects several other partitions. Looking at the stability optimisation and Reichardt et al's method compared to Arenas et al's method, the two former tend to stabilise at intermediate partitions of a size 1 larger than those detected by the latter. By using the resistance parameter $r$, Arenas et al's method alters the impact of edge weights across the network. In this instance this blends the central node into the small central community. However when considering the partition in 25 communities, this central node has 4 connections with all communities, including its own. By adding it to any community, this community would gain 4 edges pointing inside and 80 pointing outside. Therefore it is ambiguous whether this node should belong to the small central community or any of the others. This is handled in our method by keeping this node in a separate community until it is clear that it belongs to the central community of the 5 communities partition (it then shares 20 edges inside its own community and 16 edges with each of the 4 others).

On Figure 3(b), the intended partitions in 16 and then 4 communities are clearly detected. As expected the most stable partition is the partition in 4 communities, as indicated by the stability optimisation methods with the long stretch of time settling on this partition compared to the shorter plateau for 16 communities.

Considering Figure 3(c), our algorithm quickly settles on the 2 expected partitions (visualisations of the found partition confirm that this is the same as Figure 2(a)), found by Arenas et al's for about $r - r_0 \in [0.7, 2]$ and by Reichardt et al's for about $\gamma \in [0.4, 0.8]$. It also consistently settles beforehand on the partition in 4 communities, revealing the relevance of this partition, as suggested by other analysis [29]. Ronhovde and Nussinov's method mainly misses these communities.

Regarding Figure 3(d), our algorithm settles on 2 partitions, as expected from the results [27] that analysed the dataset using modularity. Arenas et al's solution for $r - r_0 \in [0.7, 1.2]$ and Reichardt et al's solution for $\gamma \in [0.2, 0.5]$ corresponds to the partition of [27]. Our algorithm also stops over on 4 partitions for $t \in [0.5, 3[ \setminus 1.5$ which is another relevant division size of the network [27]. Ronhovde and Nussinov's method misses these communities.

On Figure 3(e) we can observe several scales of relevance. Based on the knowledge of the teams distribution, a community of size 12 is expected. Such partition is detected at an early time ($t = 0.3$) with our method and is the first plateau. A normalised mutual information value of 0.919 compared with the 12 known groups can be found by our method on this plateau. Most of the nodes are therefore placed into the right communities. Regarding the remaining nodes, [18] explains that some nodes do not fit in the expected classification due to a loose intra-community connectivity and more connections out of the community. Therefore other divisions can also be of relevance. While stability settles on a few plateaus the other methods tend to detect many intermediate partitions on short intervals. As the time grows communities also grow bigger and get more stable. A partition into 3 communities is consistently identified followed by a partition with 2 communities (both also detected by Ronhovde and Nussinov's method). Analysing the former we find that it reflects the geographical locations of the teams, reflecting the fact that teams located geographically closer are more likely to play one another. This partition separates the country roughly into West, South-east and North-east. Then the partition with 2 communities divides the country into West and East. Therefore the successive stable partitions reflect the organisation of the teams, first locally with the 12 communities partition and then nationally with the partitions in 3 and 2 communities (other partitions in between may also reflect smaller geographical divisions). Another analysis could consider the large and stable communities (e.g. of size 2 or 3) and sub-partition them to analyse the games distribution at a smaller geographical scale.

Analysing the network of the characters from *Les Misérables* several divisions appear. Considering stability optimisation 2 main partitions appear, as shown on Figure 3(f), while the other methods detect more partitions on short intervals. The first one consistently identified by our method contains 5 communities and the second one contains 3 communities. In the partition into 5 communities the first is a central community containing most of the main plot characters such as Valjean, Javert, Cosette, Marius or the Thenardier. The second community relates to the story of Fantine, the third one relates to Mgr Myriel, the fourth one relates to Valjean's story as a prisoner and contains other convicts. The fifth one relates to Gavroche, another main character. Considering the partition in 3 communities, the central community is merged with the fourth community (convicts, judge, etc) and the third community. The community mainly represents characters connected to Valjean at a moment of his story. The second community remains as well as the fifth one with Marius now part of it. Overall, the detection of stable partition is clearer than for the other methods. Reichardt and Bornholdt's and Arenas et al's methods uncover some partitions but it is not clear which one is most relevant. Ronhovde and Nussinov's method misses the partitions.

These results highlight the fact that different methods provide different approaches and solutions to the problem of community partitions. Consistent or *stable* partitions are used to identify relevant divisions in a network. (Arenas et al's had a similar view considering the persistence of some partitions when changing the resolution scale in their method [3].). Considering the three tuneable methods the results show that stability optimisation tends to stabilise on fewer partitions and often more consistently than the other two methods. This is useful to identify most relevant partitions and hence inform about network structure in the absence of a priori knowledge. Once a stable partition is found, it should inform accurately about the structure of the network and can be considered *relevant*. Each community could then potentially be submitted for further partitioning.

### 5.2.2 Stable Partitions Analysis

As discussed previously, the consistency of the number of communities over successive Markov times is not sufficient to guarantee stable partitions. We introduced the use of the NMI in order to inform further about the similarity of information contained within partitions found at successive time values. Figure 4 shows for the test networks the stability optimisation curve for the Markov chain model along with the NMI between partitions found successively. We can observe that for the time values where the partitions of relevance are found, the associated NMI reaches 1 or almost. In Figure 4(a) the NMI stabilises at 1 where the number of communities is set on 5 and 26 (as discussed above), thus indicating stable partitions. For the hierarchical network the results from Figure 4(b) show that the NMI supports the detection in 4 and 16 communities. Note that for the partition in 16 communities, the NMI reaches 1 for a smaller time interval than the one during which the partition size is stable at 16. This highlights the need for a precise measure such as NMI to detect accurately stable partitions. Figure 4(c) and Figure 4(d) clearly show for the karate and dolphins networks that the most stable partitions are of size 2 even though size 3 and 4 are also detected with stable partition size and NMI reaching 1. The american football network has a partition in 12 reaching a NMI very close to 1, as shown in Figure 4(e), indicating an almost stable partition (few nodes differ while the core of the partition is the same). However the most stable partitions are found with a smaller amount of communities. Regarding the characters network from *Les Misérables* the results shown in Figure 4(f) support the relevance of the partition in 5 communities identified and commented above. The NMI is indeed stable at 1 for a large time window for this partition.

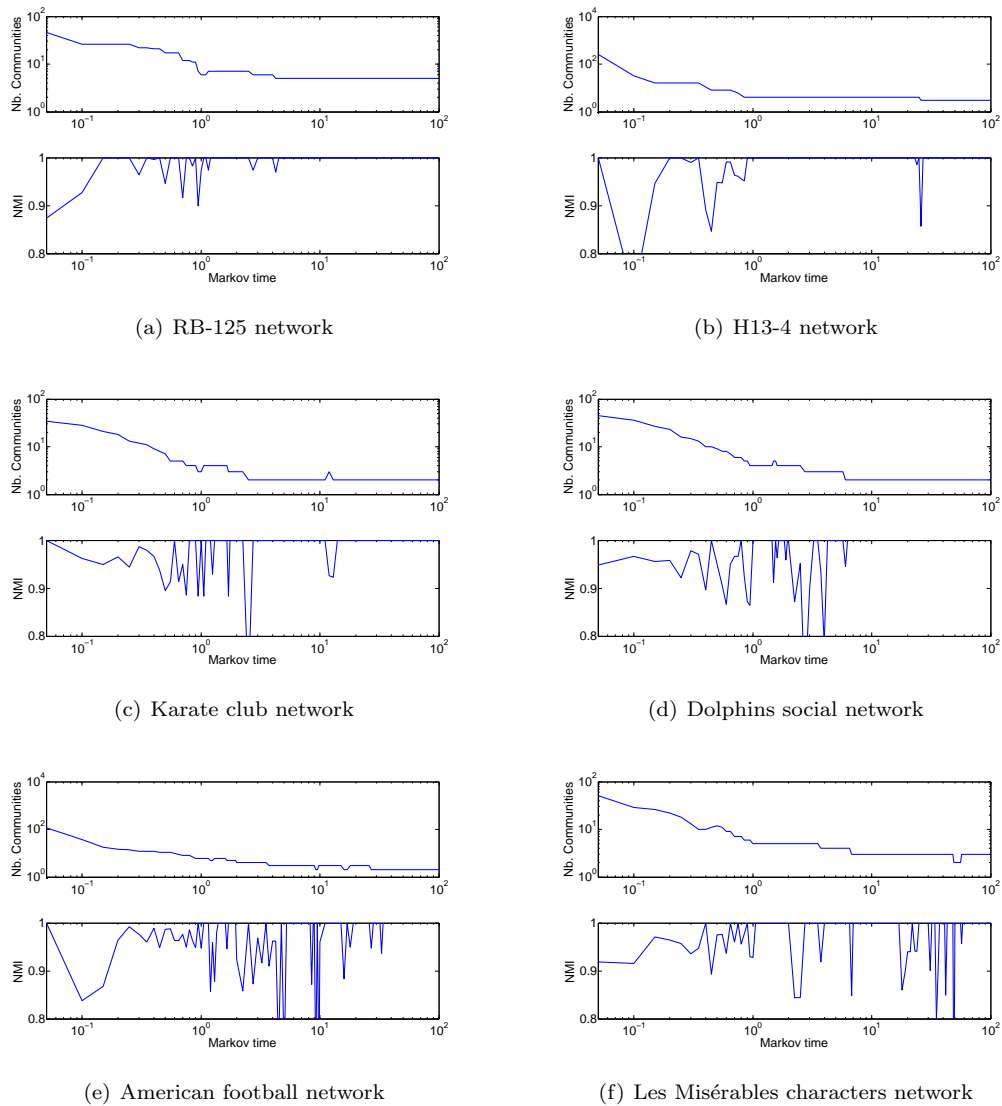

Figure 4: Number of partitions and normalised mutual information between two successive partitions for the stability optimisation methods using the discrete time (Markov chain) model.

## 5.3 Heuristics Comparison

As both the Markov chain (discrete time) and the continuous time models behave similarly we present only the results of the Markov chain model (the same results hold for the continuous time model).

### 5.3.1 Time-window Optimisation

This set of experiments compares the results obtained using a single Markov time value instead of a time-window. For the time optimised method, we use both the algorithm from Algorithm 3 and the hybrid method using the Louvain method to optimise modularity of the graph derived for each time $t$. They are compared against the initial algorithm from Algorithm 2. The results are shown in Figure 5. We can observe that the difference between the GSO runs considering the time window and the GSO runs considering only its upper bound is minimal. The curves are similar or overlapping thus suggesting that the approximation from equation (22) holds. The optimisation performed using the upper time value only combined with the Louvain method (LSO) also provides very similar results. It is noteworthy that while the LSO method may have a curve more different than the two others, the stable partitions detected are the same thus suggesting that the stability of partitions is robust to the optimisation algorithm.

### 5.3.2 Randomisation

The randomised version with and without time-optimisation are compared against the GSO. For the randomised version, each configuration was run 100 times and the results displayed are averaged over these runs. The results are shown in Figure 6.

As can be observed the randomised versions with and without time-optimisation behave similarly. Also the randomised versions provide on average results similar to the initial GSO algorithm. We can observe that in some cases, mainly regarding the synthetic networks, the average number of partitions can remain slightly above the partition sizes given by the initial GSO algorithm. Analysing the results shows that some nodes can end up isolated due to the random process. Also some small communities may not join a larger community they would be part of according to the initial GSO version's results. Taking as example the RB-125 network and considering the time up to 10, some runs where 5 communities are expected would uncover more than 5 communities. In these runs, some of the 5 five large communities (in 25 nodes) have sub communities of 5 nodes detected separately or sometimes an isolated node. The randomised method therefore provides a faster GSO algorithm but sometimes to the expense of accuracy.

A post-processing step could be used to check irregular classifications and put the isolated nodes in the neighbour community which leads to the best increase in stability. This can be done using an algorithm similar to the refinement process from [32] where each node on the edge between two communities is put in the other community to test if this move would result in a modularity increase (here it would be stability increase). This method was already an adaptation of the Kernighan-Lin algorithm [17] that aims at minimising the total edge weights across clusters by repeatedly swapping nodes belonging to different clusters that yields a maximum weight cut reduction. In our case here, one pass on selected nodes (isolated or small communities) could be sufficient to classify correctly all the isolated nodes. Therefore this post-processing step would be a minimal computational overhead. Another way, more traditional, to detect anomalies can be to perform several runs and keep for each node the classification the most consistently found over these runs.

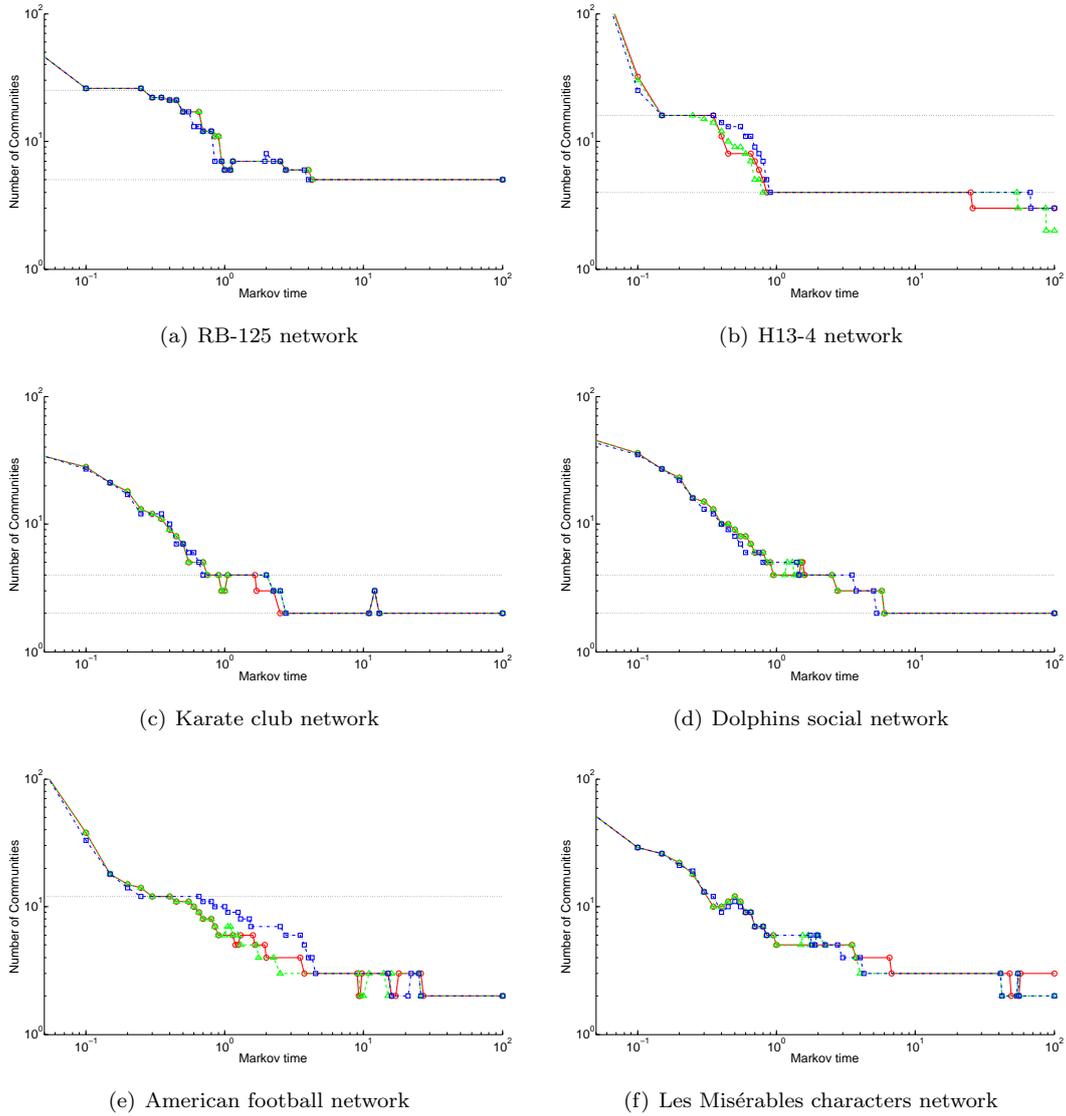

Figure 5: Number of partitions returned by the GSO and the time-optimised GSO methods (using the Markov chain model): in red and full line with circle markers is the regular GSO from Algorithm 2; in green and dashed line with square markers is the time-optimised GSO from Algorithm 3 using only one time value; in blue dotted and dashed line with triangle markers is the time-optimised version using the Louvain method (LSO).

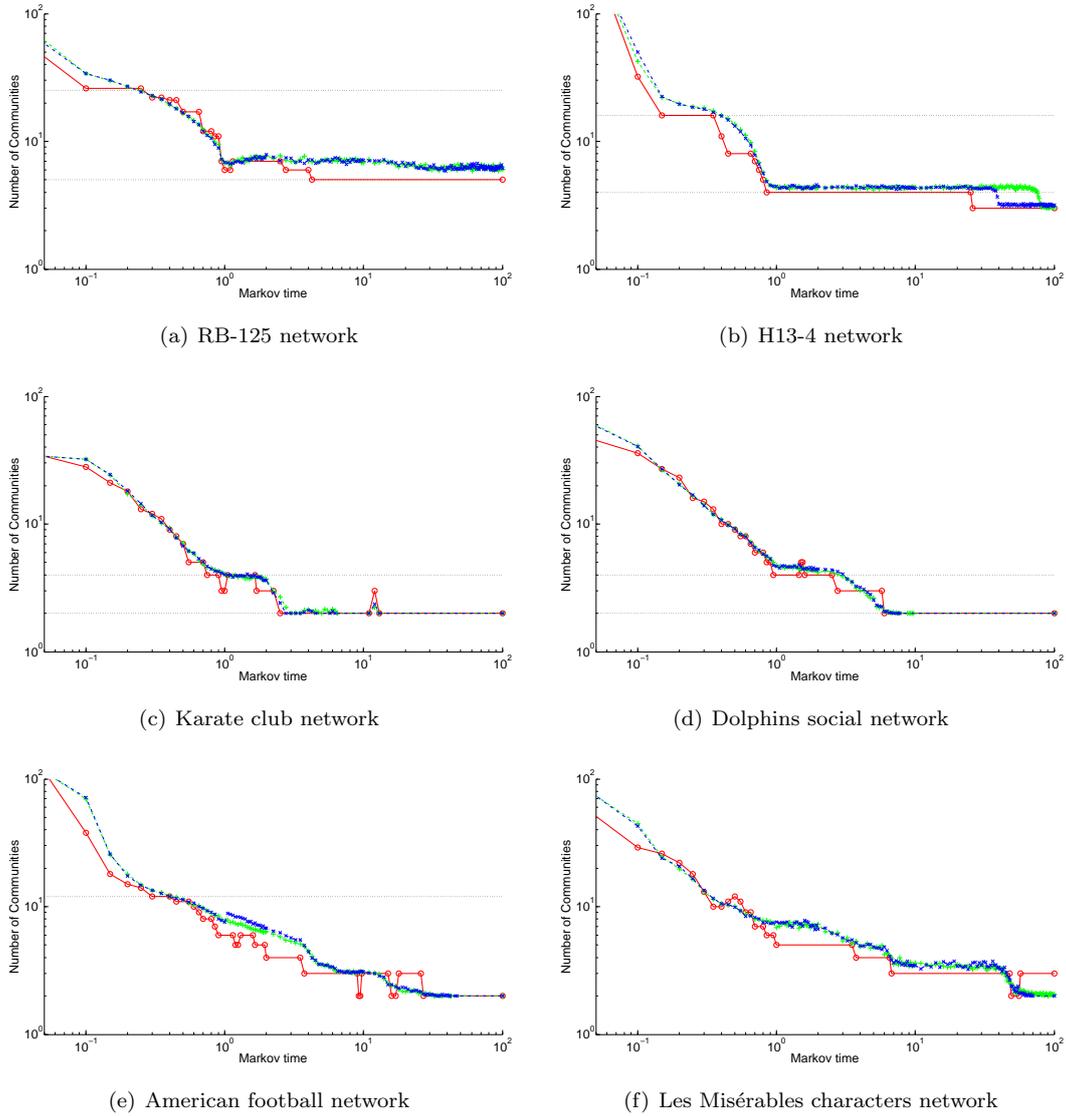

Figure 6: Number of partitions returned by the GSO and randomised GSO methods (Markov chain model): in red and full line with circle markers is the GSO from Algorithm 2; in green and dashed line with '+' markers is the randomised version; in blue dotted and dashed line with '×' markers is the randomised time-optimised version.

A pre-processing step could also be applied to the network by removing all the nodes having only one neighbour. Indeed, considering equation (18) a node with one edge that is added to the community of its only neighbour can only provide a positive $\Delta Q_S$, hence an increase in stability. Therefore these nodes can be removed at the beginning thus reducing the size of the network submitted for partitioning. Once community detection has been performed these nodes can be added again and join their neighbour's community. This pre-processing step can be used for all the techniques presented here in order to reduce the problem space.

### 5.3.3 Multi-step Aggregation

The multi-step version with values of $k \in \{2, 5, 10\}$ are compared against the GSO. The results are given in Figure 7.

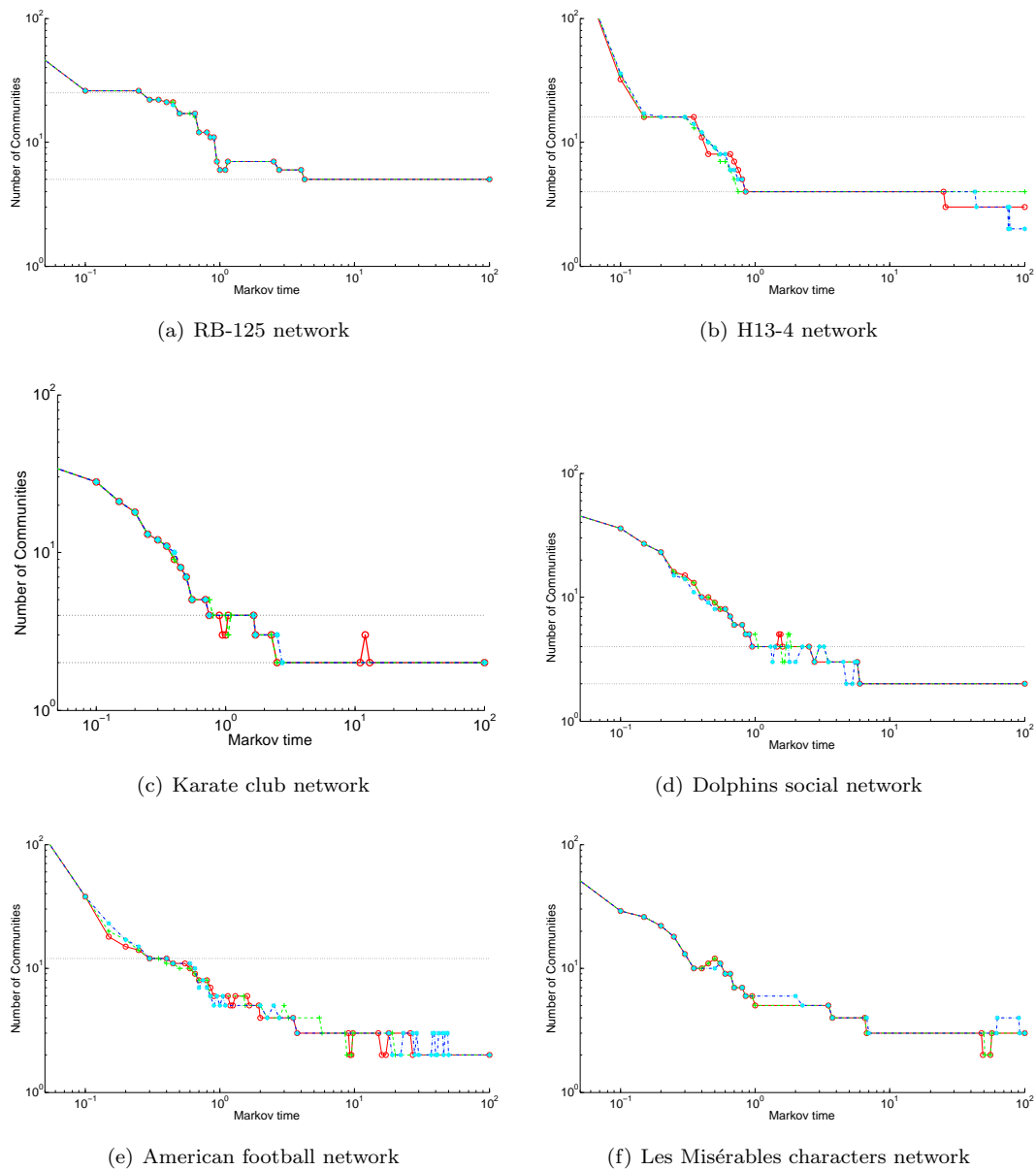

Figure 7: Number of partitions returned by the regular and multi-step GSO methods (Markov chain model). The red and full line with circle markers is the regular GSO from Algorithm 2. The remaining curves are the multi-step versions: in green and dashed line with '+' markers is with $k = 2$; in blue and dotted and dashed line with '×' markers is with $k = 5$; in cyan and dotted line with '*' marker is with $k = 10$.

From these results, we can observe that the value of $k$ has little impact on the result and especially on stable partitions. As only non-overlapping pairs are used in the algorithm, each step uses at most $k$ pairs for aggregation that do not overlap and are thus unlikely to compete against each other. Using a value of $k = 10$ is not affecting the quality of the result while speeding up the aggregation process by merging up to 10 pairs per loop. The results seem to indicate that the value of $k$ can be chosen fairly large. A generalisation with no parameter could consider at each step half (or any other subset) of the pairs and merge the ones that do not overlap.

Note that this heuristic could also be run with a time-optimised setup using only one Markov time value. It could also be combined with randomisation. All heuristics could potentially be blended together. However we cannot present all the possible combinations. We investigated several usages of all methods in order to provide concrete results and analysis of their respective behaviour. Considering those results we can expect the time-optimised version of any method to behave similarly to the GSO time-window setup. Similarly randomisation of any method will speed it up but may lead to a few isolated nodes in some runs that a post-processing step can simply detect and put in the right community.

### 5.3.4 Running Time Comparison

To illustrate the difference in running time between the various heuristics, Figure 8 provides for each method derived from the initial GSO algorithm from Algorithm 2 the average running time over 10 runs on the RB-125. The methods used are initial GSO, randomised GSO, multi-step GSO with k values taken at 2, 5, 10 and 25. All are run with and without the time-optimisation heuristic.

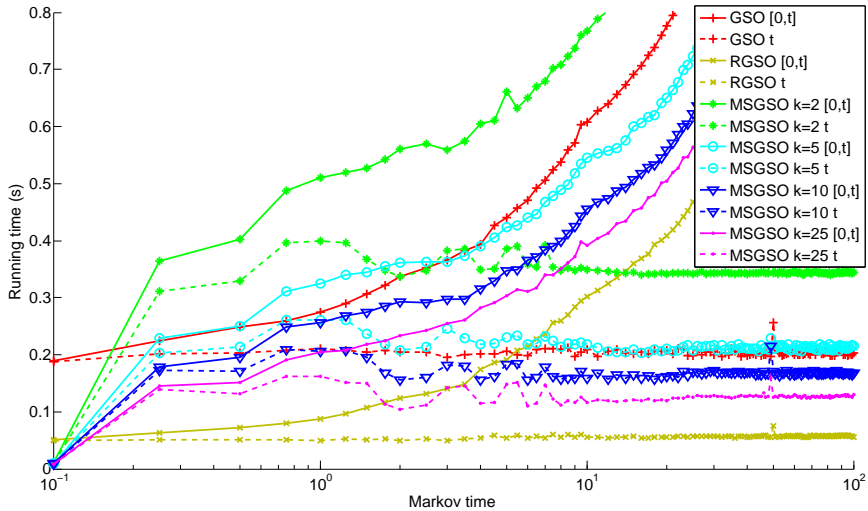

Figure 8: Running time of the various heuristics applied to the initial GSO algorithm on the RB-125 network (using the Markov chain model). Each run is performed 10 times and the time provided is the average running time of these runs.

While all the setups using a sampled time-window take longer as the Markov time grows the time-optimised versions (using one one Markov time) remain constant in execution time. Therefore the computation of the matrices $M^t$ does not affect the processing time as $t$ grows. Considering all the sampled time-window runs or all the time-optimised runs, we can observe that the randomised version is clearly the fastest of all. Regarding the multi-step models, we can observe that for values of $k$ greater than 5, the execution speed is faster than the initial GSO algorithm. The greater $k$, the faster the execution.

# 6 Detecting Overlapping Communities

## 6.1 Method

In order to apply the results of this work to overlapping communities, one can consider the line graph (or edge graph) [16] of the original network as the data to process rather than the network itself. Considering a graph $G$ with N its set of nodes and E its set of edges:

$$G = (N, E) \text{ with } E \subset N \times N$$

the line graph $L(G)$ of an undirected graph $G$ is the graph that represents the adjacencies between edges of $G$. Therefore the nodes of $L(G)$ are the edges of $G$.

$$L(G) = (E, F) \text{ with } F \subset E \times E \equiv (N \times N) \times (N \times N)$$

Two vertices of $L(G)$ are adjacent if and only if their corresponding edges are adjacent in $G$.

$$(v_1, v_2), (v_3, v_4) \in F \Leftrightarrow v_i = v_j \text{ for one } i \in 1, 2 \text{ and } j \in 3, 4$$

This idea has already been used in previous work [36, 37, 9] and is similar to the work on link communities from [1]. Indeed by working with L(G) the community detection is performed on the edges and we are thus looking at edges (or links) communities. In addition, applying this to stability optimisation enables detecting multi-scale edge communities.

## 6.2 Assessement

In order to illustrate the concept, we reuse Zachary's karate club network as it provides a social network where people can potentially belong to several communities. We also use a network of politics books from [20]. The dataset contains 105 books about US politics published around the time of the 2004 presidential election and sold by Amazon.com. The nodes represent the books and the edges between nodes represent frequent co-purchasing of books by the same buyers. A suggestion of book labels has been given in [32] according to their political tendency: liberal, neutral, or conservative. The overlapping community detection seems to be more relevant than a crisp community detection approach as books on political tendencies like many fuzzy notions may sometimes share features and thus overlap. A book can be neutral but still addressing interesting concepts for readers from one side or the other. A book can also be neutral with a tendency to one side or the other. As a result it seems that the expected communities could be the two communities: liberal and conservative, with an overlap for fairly neutral books.

The results of the overlapping community detection are presented in Figure 9. We can observe on Figure 9(a) that the most stable partitions are found for times around 1 with 4 communities, and then times after 2 with 2 communities. These results are consistent with the knowledge of the network, and with the results found with previous experiments. In addition they encompass the overlapping information between communities. Figure 10(a) and Figure 10(b) show those two partitions. On Figure 9(b) stable partitions are found between times 1 and 2 with 3 communities, and then mostly after time 7 with 2 communities. Figure 10(d) and Figure 10(e) show those two partitions.

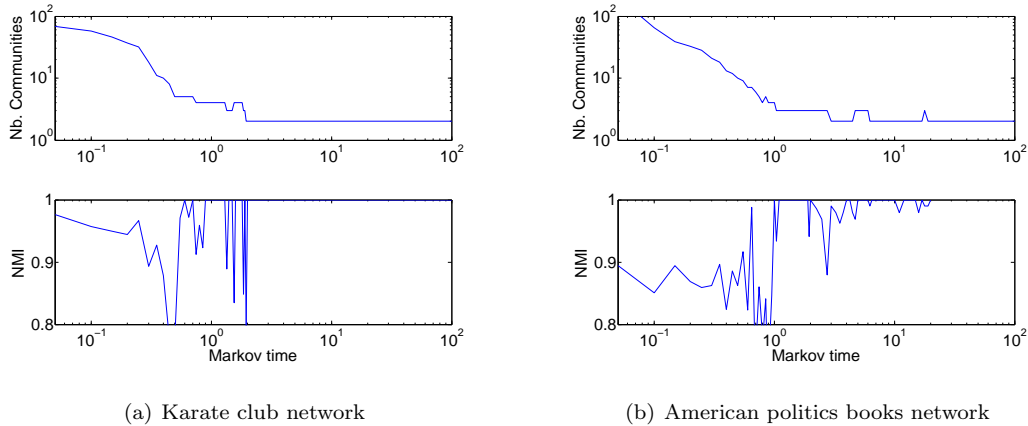

(a) Karate club network  (b) American politics books network

Figure 9: Number of partitions and normalised mutual information between two successive partitions for the stability optimisation methods applied to overlapping communities on the Karate club network and the American politics books network. The results are for the discrete time (Markov chain) model.

The neutral labels by Newman shown in red with a star marker in Figure 10(c) are scattered around the two main groups, thus suggesting that neutral books might not reflect a community in the definition accepted here but rather a set of books that can be of interest to both communities. The neutral books on the left are mainly represented by the third edge community, the ones on the right seem to be more densely linked to the blue community, and are thus classified in the blue community. The remaining neutral books can be found on the edge between communities. Using the edge communities, we can provide a classification of books with overlap to indicate shared interests as well as a potential third edge community. The third small community (in red) may reveal a subset of neutral books sharing some features that other neutral books do not, or books belonging to the green community but still neutral enough to be of interest to the blue community. The overall classification we obtain is not exactly the same as the one suggested by Newman but shares strong similarities. These divisions are perfectly sensible with the set being analysed. We can thus verify the hypothesis formulated above about the distribution of books within communities.

One asset of this approach is that it is compatible with all the crisp boundaries community detection methods and hence all the work discussed previously above can be used here. One criticism of the approach is that it always finds overlapping node communities on the boundaries between communities as by definition a node that joins two edges belonging to two different edge communities belongs to two associated node communities. While the overlapping between communities seems to be a common feature of social networks, it might not always be the case for all kinds of networks.

# 7 Conclusion

Stability optimisation is a technique aiming at optimising a partition quality measure called stability [8]. This technique was introduced in [24] in its basic form. However further heuristics, optimisations and applications had not been assessed. This work presented a broader investiga-tion of stability as an optimisation criterion for a greedy approach.

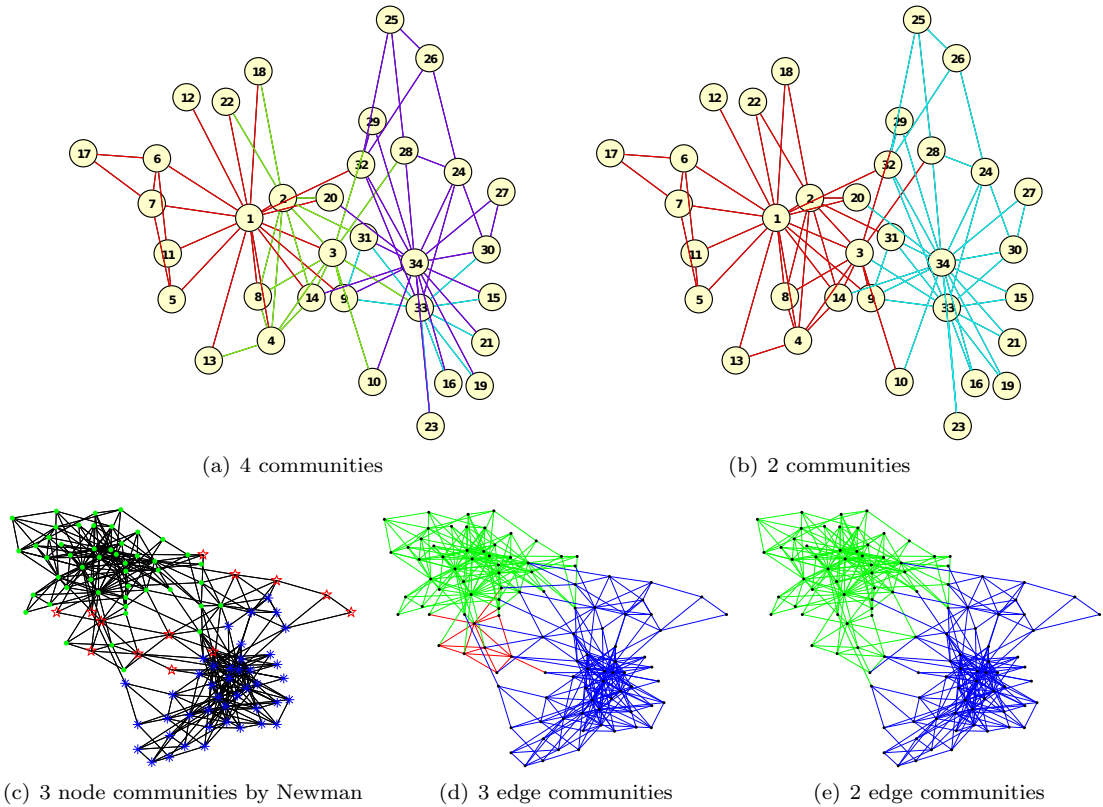

Figure 10: (a) and (b): Partitions in 4 and 2 edge communities on Zachary's karate club network. (c): Partitions of the American politics books network in 3 node communities by Newman; (d) and (e): Partition of the American politics books network into 3 and 2 edge communities.

Stability is a measure that encompasses other measures such as the well known modularity [33]. Its optimisation can be achieved similarly to modularity optimisation but enables accurate multi-scale community detection by the use of Markov time as a resolution parameter. The results showed that stability optimisation accurately detects partitions of relevance by uncovering stable partitions. Several heuristics have been devised and tested to provide speed improvement and alternative ways to build the communities. The first optimisation is based on the reduction of the number of Markov time values used for stability computation. Experiments showed that this heuristic can provide significant gain in speed with no loss of accuracy. The second optimisation is based on a randomisation of the algorithm. Experiments showed that this heuristic provides significant speed performance increase with no significant loss of accuracy. The third heuristic is a multi-step version of the algorithm. Experiments showed that this heuristic can also provide significant gain in speed. It is noteworthy that these three heuristics can also be combined together. Then stability optimisation was also combined with the Louvain method to show that other modularity optimisation methods can be combined with stability optimisation with no overhead. This combination also enabled to study further the robustness of the uncovered partitions, not only to the Markov time, but also to the aggregation algorithm used. Experiments overall demonstrated that stable partitions are uncovered by all stability optimisation methods consistently.

The results showed that multiple levels of organisations are clearly identified when optimising stability over time. Stability optimisation tends to settle for longer on fewer partitions than other related approaches considered here, thus highlighting better partitions of relevance. Stability optimisation also converges towards large and *stable* clusters. This behaviour differs from those of other approaches and in the absence of a priori knowledge our method has therefore the advantage of leading to stable and relevant communities from where a deeper analysis could be performed in each community subgraph.

Finally stability optimisation was also applied to the detection of overlapping communities by using the line graph of the initial graph [16]. This approach had been used in [36, 37, 9]. However this work enabled the detection of stable edge (or link) communities, thus offering not one but several partitions of relevance based on the uncovered scales.

The method was tested on binary undirected graphs as it was deemed appropriate for the purpose of this work but it can equally be applied to directed weighted graphs. Indeed even if the initial adjacency matrix contains only zeros and ones the adjacency matrices $A_t$ used for stability optimisation are real-valued matrices.